\documentclass[12pt]{article}
\usepackage[margin=2.5cm]{geometry}
\usepackage{amsmath,amssymb,amsfonts}
\usepackage{mathrsfs}
\usepackage{graphicx}
\usepackage{tikz}
\usepackage{hyperref}
\definecolor{freezepink}{RGB}{219,68,150}

\newif\ifhighlight \highlightfalse
\newif\ifhighlightfirst \highlightfirstfalse
\definecolor{revised}{RGB}{0,130,80}
\definecolor{revisedb}{RGB}{0,85,212}
\DeclareRobustCommand{\chg}[1]{\ifhighlight\ifhighlightfirst\textcolor{revised}{#1}\else#1\fi\else#1\fi}
\DeclareRobustCommand{\chgb}[1]{\ifhighlight\textcolor{revisedb}{#1}\else#1\fi}
\ifhighlight\ifhighlightfirst\tikzset{chg/.style={revised}}\else\tikzset{chg/.style={}}\fi\else\tikzset{chg/.style={}}\fi
\providecommand{\doi}[1]{\chg{\href{https://doi.org/#1}{doi:\nolinkurl{#1}}}}

\title{Variational principles for dissipative continua: transport theorems, inertia, and the minimum--maximum duality}
\author{Wolfgang Flachberger$^{1,*}$, Lukas Einkemmer$^{1}$ \\[6pt]
\small $^1$Department of Mathematics, University of Innsbruck, 6020 Innsbruck, Austria \\
\small $^*$Corresponding author: wolfgang.flachberger@uibk.ac.at}
\date{}

\begin{document}
\maketitle

\begin{abstract}
Variational principles for dissipative processes come in two families: \textit{minimum} principles at nonequilibrium stationary states (Helmholtz, Kirchhoff, Prigogine) and \textit{maximum} principles for relaxation in continuum mechanics (Ziegler, Hackl, Fischer \cite{ziegler1963, hackl2008}). We contribute to the understanding of the thermodynamic extremal principle (TEP) by investigating its extremal properties\chg{, and uncover} the connection between the two families\chg{, whose distinction was recently made precise for diffusion \cite{svoboda2023note}}: the extremal character is decided by the regime, which is either relaxation \chg{at frozen state, as} in a closed system (maximum dissipation)\chg{,} or a driven stationary state in an open system (minimum dissipation), while the underlying Rayleighian functional is convex and minimized in both cases. We further examine the Reynolds transport theorem, a useful tool for fixing the energy rates of continua, and the role of inertia, recently argued to lie beyond the reach of Rayleighian principles \cite{beris2024}. With inertia the driving force must be held fixed during the variation, which is less elegant than in the inertialess case, where it is a state function. \chg{Writing the energy rate as rates times driving forces} still yields the correct equations, and an extremal statement survives whenever the frozen force does not depend on the varied rate. This holds directly for unsteady Stokes flow, and for Navier--Stokes flow after a change to material coordinates, in which the particle acceleration is the second time derivative of the state and the convective term, \chg{nonvariational} in the Eulerian description, does not appear. The framework is tested on the Navier--Stokes family (Stokes, unsteady Stokes\chg{,} and Navier--Stokes flow, in order of increasing complexity) and on dissipative Lagrangian mechanics.
\end{abstract}

\section{Introduction}

Variational principles are a cornerstone of physics: Fermat's least time in optics, \chg{Hamilton's} stationary action in mechanics, and the Einstein--Hilbert action of general relativity all condense a physical law into the stationarity of a single scalar functional\chg{, and Feynman's path integral recovers this stationarity in the classical limit}. Over the past decade these principles have also proven \chg{valuable} for developing numerical methods for physical phenomena \cite{flachberger2024, beronvera2021}: a variational problem is typically the most compact representation of a coupled multiphysics problem, and can be used to derive the \chg{weak} form automatically for use with, e.g., the finite element method. For dissipative systems the picture is less settled \cite{martyushev2006}, and two questions motivate the present work.

First, the extremal principles divide into two families that reflect the different scientific communities from which they arose. \textit{Minimum} principles \chg{apply to} nonequilibrium stationary states\chg{:} Helmholtz's minimum dissipation for Stokes flow \cite{helmholtz1868}, Kirchhoff's for stationary currents \cite{kirchhoff1848}, \chg{and} Prigogine's minimum entropy production \cite{kondepudi2014modern}\chg{. They} belong to thermodynamicists and chemists (\chg{Prigogine a chemist} by training, Helmholtz a physician-turned-physicist), whose paradigm is the driven open system of chemistry and chemical engineering: reactants in, products out, entropy production minimized at the stationary state in between (Section~\ref{sec:driven}). \textit{Maximum} principles instead \chg{apply to} relaxation in continuum mechanics and materials science\chg{:} Ziegler's maximum rate of dissipation \cite{ziegler1963}, the principle of maximum dissipation \chg{in} inelasticity, whose relation to dissipation potentials was clarified by Hackl and Fischer \cite{hackl2008}, and the thermodynamic extremal principle (TEP) of materials science \cite{svoboda2005application, fischer2014review}\chg{. They} belong to engineers and materials scientists, whose paradigm is inelastic material response \chg{(}plasticity, creep, phase transformation\chg{)} in a closed system releasing stored free energy at frozen state through rates that maximize the dissipation (Section~\ref{sec:relaxation}). \chg{Onsager's principle of least dissipation of energy \cite{onsager1931reciprocal} is, despite its name, not a stationary-state principle: it is an instantaneous principle for the rates, read as the maximum principle by Svoboda et al.\ \cite{svoboda2023note}, and \chgb{the functional it minimizes} is the Rayleighian functional that connects the two families (Section~\ref{sec:heritage}).} That instantaneous maximization and stationary-state minimization of the entropy production are distinct statements has been noted before \cite{struchtrup1998, polettini2013}, and for multicomponent diffusion the distinction has recently been made precise \cite{svoboda2023note}: forces given and rates varied at an \chgb{instant versus} fields varied under fixed boundary data at the \chg{stationary} state.

Second, inertia sits uneasily with these principles\chg{.} Beris and Edwards \cite{beris2024} recently pointed out that Rayleighian extremum principles remain restricted to cases where the kinetic energy is negligible\chg{,} yet fluid mechanics routinely requires viscosity and inertia together. Hackl, Svoboda\chg{,} and Fischer \cite{hackl2024coupling} \chgb{proposed a} coupling of the TEP with Hamilton's principle in exactly this context\chgb{, in which the Lagrangian is extended by a dissipation-force term; the present work follows a different route through the rate of the total energy (Section~\ref{sec:lagrangian})}.

\begin{figure}[ht]
\centering
\includegraphics[width=\textwidth]{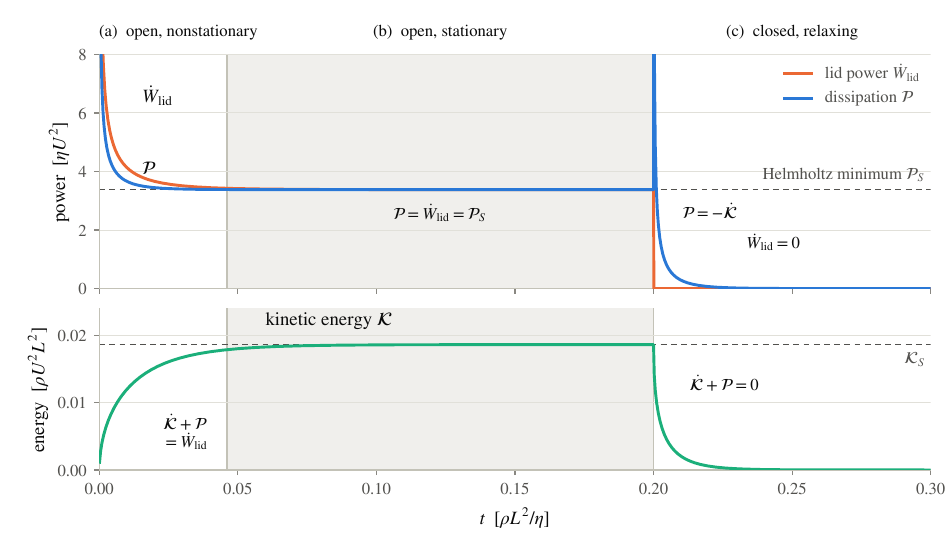}
\caption{\chgb{Unsteady Stokes flow in a lid-driven square cavity ($\rho = \eta = L = U = 1$; lid velocity $U\,16x^2(1-x)^2$; Taylor--Hood elements on a $64 \times 64$ mesh; backward Euler with $\Delta t = 1.25 \times 10^{-4}$). Top: dissipation $\mathcal{P}$ and lid power $\dot{W}_{\mathrm{lid}}$; bottom: kinetic energy $\mathcal{K}$; dashed lines: the stationary values $\mathcal{P}_S$ and $\mathcal{K}_S$. Phase (a), lid started from rest: $\dot{\mathcal{K}} + \mathcal{P} = \dot{W}_{\mathrm{lid}}$, and $\mathcal{P}$ decreases monotonically to Helmholtz's minimum $\mathcal{P}_S = 3.379\,\eta U^2$, reached within $0.1\%$ at $t = 0.046$. Phase (b), stationary state: $\mathcal{P} = \dot{W}_{\mathrm{lid}} = \mathcal{P}_S$. Phase (c), lid stopped at $t = 0.2$: $\dot{\mathcal{K}} + \mathcal{P} = 0$, and the decay rate approaches twice the smallest Stokes eigenvalue, $2\lambda_1 \approx 105\,\eta/(\rho L^2)$. The curves leave the frame at the impulsive start and stop, where the dissipation in the wall boundary layer diverges like $t^{-1/2}$. The three phases are interpreted in Section~\ref{sec:minmax}.}}
\label{fig:cavity}
\end{figure}

\chgb{Both questions meet in a classical benchmark of fluid mechanics, the flow in a lid-driven square cavity \cite{shankar2000}, here computed with the unsteady Stokes equation, which neglects convection but keeps inertia through the time derivative. Figure~\ref{fig:cavity} shows one computation in three phases. In phase (a) the lid is started from rest, and the dissipation decreases monotonically. In phase (b) the flow has become stationary: the power supplied by the lid is dissipated entirely, and by Helmholtz's theorem this stationary dissipation is the smallest among all incompressible flows with the same lid velocity. In phase (c) the lid is stopped, and the stored kinetic energy is dissipated until the fluid comes to rest. The same equation thus produces a state of minimum dissipation and, after the lid stops, a decay that proceeds, as Section~\ref{sec:relaxation} shows, through the rates of maximal dissipation. Explaining why one functional yields both, with inertia present, is the subject of this work.}

The present work is a contribution to the understanding of the TEP. We investigate its extremal properties and find the two families connected through a single convex functional, the Rayleighian, whose minimum \chgb{appears} as Ziegler's maximum during relaxation and as the Helmholtz--Prigogine minimum at a driven stationary state (Section~\ref{sec:minmax})\chg{. Svoboda et al.\ \cite{svoboda2023note} established the maximum and the minimum as two different extremization tasks for multicomponent diffusion; we extend this distinction to fluids and to systems with inertia and show that both tasks are \chgb{two sides} of one Rayleighian minimum}. We further investigate the role of the Reynolds transport theorem, which turns out to be a useful tool: the choice of control volume fixes the energy rate the TEP \chgb{requires and} with it the structure of the resulting equations. The TEP is elegant for inertialess systems, where the driving forces are state functions. With inertia the driving force contains the acceleration and must be frozen during the variation, which is less natural. The same rule applies in Navier--Stokes flow and in Lagrangian mechanics, and \chg{writing $\dot{\mathcal{F}}$ as rates times driving forces} still delivers the correct equations, so that the TEP also covers dynamics that are not overdamped. An extremal statement survives whenever the frozen driving force does not depend explicitly on the varied rate: for unsteady Stokes flow and constant-mass mechanics the minimum persists directly, and for Navier--Stokes flow it is recovered by a change of perspective to material coordinates, in which the particle acceleration is the second time derivative of the state; for configuration-dependent kinetic metrics only stationarity remains in the given coordinates. The argument is carried by the Navier--Stokes family of Section~\ref{sec:nsfamily}, that is, by Stokes, unsteady Stokes\chg{,} and Navier--Stokes flow in order of increasing complexity, and by dissipative Lagrangian mechanics (Section~\ref{sec:lagrangian}). Diffusion is deliberately left aside: there the TEP works \chg{well and has become an established tool}, especially for multicomponent diffusion, where its use has been systematized in \cite{flachberger2024}; the scope here is fluids and mechanics, where inertia enters. Throughout, the dissipation is quadratic in the rates; Section~\ref{sec:linearity} states what this restricts and shows that homogeneity \chgb{of the dissipation in the rates already suffices for both extremal statements}.

\section{The Thermodynamic Extremal Principle}
\label{sec:tep}

\subsection{Heritage}
\label{sec:heritage}

The idea that dissipative dynamics follows from an extremum principle predates irreversible thermodynamics: Kirchhoff proved in 1848 that stationary electric currents minimize the generated heat \cite{kirchhoff1848}, and Rayleigh \cite{rayleigh1873} introduced the dissipation function~--- one half of the rate of energy dissipation~--- as a velocity potential for friction forces. Onsager \cite{onsager1931reciprocal} elevated this to a general principle of \textit{least dissipation of energy} near equilibrium, and Prigogine established minimum entropy production at nonequilibrium stationary states \cite{kondepudi2014modern}. Ziegler, by contrast, formulated a principle of \textit{maximum} rate of dissipation for systems relaxing at a given state \cite{ziegler1963}, continued in solid mechanics by Hackl and Fischer, who established the relation between the principle of maximum dissipation and evolution laws derived from dissipation potentials \cite{hackl2008}, and systematized in materials science as the thermodynamic extremal principle \cite{svoboda2005application, fischer2014review, emmerich2003diffuse} (see \cite{martyushev2006} for a review). In soft matter it appears as the Onsager variational principle \cite{doi2011}, \chgb{which minimizes the functional used here, the Rayleighian}.

A remark on nomenclature is in order, since this lineage has left many names for closely related objects. Our naming follows the pattern set by mechanics: script letters denote functions, such as the Lagrangian $\mathscr{L}$ and the Hamiltonian $\mathscr{H}$, while calligraphic letters denote functionals, such as the \textit{Rayleighian functional} $\mathcal{R} = \int_V \mathscr{R}\, dV$ \cite{doi2011, beris2024}. For the principle we retain \chgb{the name} \textit{thermodynamic extremal principle}\chgb{. The names \textit{least dissipation of energy} and \textit{maximum} or \textit{minimum entropy production} already say whether the dissipation is minimized or maximized, and the name \textit{Onsager variational principle} does not refer to an extremum at all. \textit{Thermodynamic extremal principle} asserts an extremum without fixing its kind, which matches our results: the Rayleighian is minimized in every regime, while the dissipation is maximized, minimized, or neither, depending on the regime. The GENERIC formalism (General Equation for the Non-Equilibrium Reversible--Irreversible Coupling)} and the bracket formalisms \cite{grmela1997, beris2024} describe the \textit{complete} dynamics, reversible and irreversible together; the TEP characterizes the dissipative part alone, and the recurring rule that reversible currents enter the variation as frozen data ($D\boldsymbol{u}/Dt$ in Section~\ref{sec:ns}, $\tfrac{d}{dt}\partial \mathscr{L}/\partial \dot{q}$ in Section~\ref{sec:lagrangian}) expresses their degeneracy condition: reversible dynamics produces no dissipation.

\subsection{Derivation from constrained stationarity of the dissipation}
\label{sec:derivation}

Consider a system with free energy $\mathcal{F}$ depending on a set of state variables, and dissipative mechanisms described by generalized rates $\dot{q}_i$. Near equilibrium the rate of dissipation $\mathcal{P}$ is a positive-definite quadratic form in the rates:
\begin{align}
    \mathcal{P}(\dot{q}_1, \dot{q}_2, \ldots) = \sum_{ij} (L^{-1})_{ij}\, \dot{q}_i \dot{q}_j \geq 0, \qquad L_{ij} = L_{ji}\chg{,} \label{eq:entropy_prod}
\end{align}
where $L_{ij}$ is the matrix of Onsager coefficients, positive definite and, by Onsager's reciprocal relations \cite{onsager1931reciprocal}, symmetric; its inverse $(L^{-1})_{ij}$ is the kinetic resistance. The coefficients may depend on the state, which is frozen in the variation; convex dissipations that are homogeneous of degree $k > 1$ but not quadratic in the rates are treated in Section~\ref{sec:linearity}. At constant temperature $T$, the dissipation is proportional to the rate of internal entropy production, $\mathcal{P} = \chg{T\, d_i S/dt}$; we work with $\mathcal{P}$, which has units of power, so that it can be balanced directly against the free energy rate. The rate of free energy change $\dot{\mathcal{F}}$ is linear in the rates:
\begin{align}
    \dot{\mathcal{F}} = \sum_i X_i \dot{q}_i\chg{,} \label{eq:F_rate}
\end{align}
where $X_i = \partial \mathcal{F} / \partial q_i$ is evaluated at the frozen state, so that \eqref{eq:F_rate} is \chgb{simply} the chain rule. With this convention $X_i$ carries the sign opposite to the driving force $F_i = -\partial\mathcal{F}/\partial q_i$ commonly used in the literature, hence the minus sign in \eqref{eq:onsager_law} below. If the system exchanges \chg{no work and no matter with its surroundings and is held at constant temperature by a heat bath}, the free energy decreases exactly at the rate at which energy is dissipated:
\begin{align}
    \dot{\mathcal{F}} + \mathcal{P} = 0\chg{.} \label{eq:energy_balance}
\end{align}
This balance \chgb{expresses energy conservation}; the second law enters through $\mathcal{P} \geq 0$, which guarantees $\dot{\mathcal{F}} \leq 0$.

The TEP selects, among all rates compatible with the balance \eqref{eq:energy_balance}, those that render the dissipation \textit{stationary}. Using a Lagrange multiplier $\lambda$:
\begin{align}
    \delta_{\dot{q}_i,\lambda} \left[ \mathcal{P} + \lambda \big( \mathcal{P} + \dot{\mathcal{F}} \big) \right] = 0\chg{.} \label{eq:tep_constrained}
\end{align}
The stationarity conditions with respect to $\dot{q}_i$ read:
\begin{align}
    2(1+\lambda) \sum_j (L^{-1})_{ij}\, \dot{q}_j + \lambda X_i = 0 \qquad \forall\, i\chgb{.} \label{eq:stationarity}
\end{align}
\chgb{For $\lambda \neq -1$, this system is solved by} $\dot{q}_i = -\frac{\lambda}{2(1+\lambda)} \sum_j L_{ij} X_j$. Inserting this \chgb{solution} into the constraint \eqref{eq:energy_balance} yields:
\begin{align}
    \dot{\mathcal{F}} + \mathcal{P} = -\frac{\lambda(2+\lambda)}{4(1+\lambda)^2} \sum_{ij} L_{ij} X_i X_j = 0
    \qquad \Longrightarrow \qquad \lambda \in \{0,\, -2\}\chg{.} \label{eq:lambda}
\end{align}
Away from equilibrium ($X \neq 0$), the multiplier equation thus has \textit{two} roots, and both matter. The root $\lambda = 0$ gives the trivial rest state $\dot{q}_i = 0$: nothing moves, nothing is dissipated, and the balance is satisfied with $\dot{\mathcal{F}} = \mathcal{P} = 0$. The root $\lambda = -2$ gives the Onsager evolution law:
\begin{align}
    \dot{q}_i = -\sum_j L_{ij}\, X_j\chg{.} \label{eq:onsager_law}
\end{align}

The nature of these two stationary points is revealed by the geometry of the balance \eqref{eq:energy_balance}: the admissible rates form an ellipsoid through the origin. The dissipation $\mathcal{P}$ attains its constrained \textit{minimum} $\mathcal{P} = 0$ at the origin itself (the trivial root $\lambda = 0$) and its constrained \textit{maximum} at the antipodal point, which is precisely the Onsager law \eqref{eq:onsager_law} ($\lambda = -2$). Among all rates that could account for the free energy released per unit time, the system thus selects those that dissipate the most~--- it descends the free energy landscape as steeply as the kinetic resistances permit. This is the extremal statement of the Ziegler family \cite{ziegler1963, hackl2008} \chg{(for two fluxes, the ellipse is drawn in \cite[Fig.~1]{svoboda2023note})}, and it belongs to the \textit{relaxation} regime, in which free energy is released at frozen state.

Substituting $\lambda = -2$ into the augmented functional of \eqref{eq:tep_constrained} gives $\mathcal{P} - 2\big(\mathcal{P} + \dot{\mathcal{F}}\big) = -2\big(\dot{\mathcal{F}} + \tfrac{1}{2}\mathcal{P}\big)$, so that up to the overall factor $-2$ the constrained problem is equivalent to the \textit{unconstrained} stationarity of the functional
\begin{align}
    \mathcal{R}[\dot{q}_1, \dot{q}_2, \ldots] = \dot{\mathcal{F}} + \tfrac{1}{2}\mathcal{P}\chg{,} \label{eq:tep_general}
\end{align}
known as the Rayleighian functional \cite{rayleigh1873, doi2011}. The negative prefactor means only stationarity carries over; the extremal character must be assessed on $\mathcal{R}$ itself, and turns out to be the opposite: $\mathcal{R}$ is the sum of a linear and a positive-definite quadratic form, hence strictly convex, so its unconstrained stationary point is a true \textit{minimum}~--- once again the Onsager law \eqref{eq:onsager_law}. We thus arrive at the central structural fact of the TEP, on which Section~\ref{sec:minmax} builds:
\begin{quote}
    The physical rates \eqref{eq:onsager_law} are simultaneously the unconstrained \textbf{minimizer} of the Rayleighian functional $\mathcal{R} = \dot{\mathcal{F}} + \tfrac{1}{2}\mathcal{P}$ and the \textbf{maximizer} of the dissipation $\mathcal{P}$ over all rates satisfying the energy balance $\dot{\mathcal{F}} + \mathcal{P} = 0$.
\end{quote}

\chg{For the two-flux example of \cite[Fig.~1]{svoboda2023note}, with $\dot{\mathcal{F}} = -2\dot{q}_1 - 3\dot{q}_2$ and $\mathcal{P} = 3\dot{q}_1^2 + 3\dot{q}_1\dot{q}_2 + 2\dot{q}_2^2$, the unconstrained minimum of $\mathcal{R}$ lies at $(\dot{q}_1, \dot{q}_2) = (-2/15,\, 8/5)$, which is precisely the constrained maximum of the dissipation\chgb{, shown as a black dot in that figure}.}

The complementary regime is the \textit{nonequilibrium stationary state}, where $\dot{\mathcal{F}} = 0$ while $\mathcal{P} > 0$: dissipation is sustained by external throughput while no stored free energy is consumed, and the balance \eqref{eq:energy_balance} can no longer be imposed as a constraint coupling the rates to frozen forces. The multiplier term of \eqref{eq:tep_constrained} drops, and the problem reduces to requiring stationarity of $\mathcal{P}$. This regime is also captured by the Rayleighian \eqref{eq:tep_general}. By setting $\dot{\mathcal{F}} = 0$ the Rayleighian simplifies to:
\begin{align}
    \mathcal{R} = \tfrac{1}{2} \mathcal{P}\chg{.} \label{eq:tep_stationary}
\end{align}
Since $\mathcal{P}$ is convex, this stationary point is a \textit{minimum}~--- the extremal statement of the \chg{Helmholtz--Prigogine} family \cite{helmholtz1868, kirchhoff1848}. Throughout this work, then, the TEP is identified with the stationarity of the Rayleighian functional, which generates both families: whether it \chgb{appears} as Ziegler's maximum or as the Helmholtz--Prigogine minimum is decided by the regime. Section~\ref{sec:minmax} develops the two side by side.

\subsection{Distinction from Hamilton's principle}
\label{sec:distinction}

Despite sharing the vocabulary of Lagrangians and Euler--Lagrange equations, the TEP is \chg{\textit{fundamentally different}} from Hamilton's principle, in three respects. \textit{What is varied}: Hamilton's principle varies the trajectory $q(t)$ over an interval $[t_1,t_2]$, with $\delta q$ and the induced $\delta \dot{q} = \frac{d}{dt}(\delta q)$ coupled, so that $q$ and $\dot q$ are determined together; the TEP instead varies the rates $\dot{q}_i$ independently, at a single instant, for a given (frozen) state~--- only the rates are unknown. \textit{Euler--Lagrange equations}: Hamilton's principle yields temporal equations $\frac{\partial \mathscr{L}}{\partial q} - \frac{d}{dt}\frac{\partial \mathscr{L}}{\partial \dot{q}} = 0$; the TEP yields algebraic (or spatial) relations between the rates, with no temporal differentiation of the stationarity conditions. \textit{Extremal character}: Hamilton's action is stationary but generally neither minimal nor maximal, whereas the Rayleighian functional is strictly convex in the rates and attains a true minimum, \chgb{while} the dissipation itself \chgb{is} maximized over the \chg{balance-consistent} rates (Section~\ref{sec:derivation})\chg{. For} Stokes and unsteady Stokes flow the corresponding minimum statements hold on each problem's constraint manifold (Sections~\ref{sec:stokes}, \ref{sec:ustokes}\chg{,} and~\ref{sec:minmax}), and for Navier--Stokes flow in material coordinates (Section~\ref{sec:ns})\chg{. With} a configuration-dependent kinetic metric only stationarity survives in the given coordinates (Section~\ref{sec:lagrangian}).

%Confusing the two frameworks~--- for instance, by applying temporal Euler--Lagrange equations to the Rayleighian functional~--- leads to incorrect evolution equations, as was recently demonstrated for vacancy diffusion \cite{flachberger_corrected}.

\subsection{The Reynolds transport theorem}
\label{sec:rtt}

Formulating the TEP for continua requires care with the free energy rate $\dot{\mathcal{F}}$. For a system occupying a volume $V$ with free energy density $f(\boldsymbol{x},t)$, the total free energy is $\mathcal{F} = \int_V f\, dV$, and its rate depends critically on whether $V$ is fixed in space or moves with the material. The Reynolds transport theorem provides the rigorous connection:

\textbf{Fixed control volume:}
\begin{align}
    \frac{d}{dt}\int_V f \, dV = \int_V \frac{\partial f}{\partial t} \, dV\chg{.} \label{eq:rtt_fixed}
\end{align}

\textbf{Material (moving) control volume:}
\begin{align}
    \frac{d}{dt}\int_{V(t)} f \, dV = \int_{V(t)} \left( \frac{\partial f}{\partial t} + \nabla \cdot (f \chg{\boldsymbol{u}}) \right) dV = \int_{V(t)} \frac{Df}{Dt} \, dV + \int_{V(t)} f \, \nabla \cdot \chg{\boldsymbol{u}} \, dV\chg{,} \label{eq:rtt_material}
\end{align}
where $\chg{\boldsymbol{u}}$ is the velocity of the material and $D/Dt = \partial/\partial t + \chg{\boldsymbol{u}} \cdot \nabla$ is the material derivative.

The choice between \eqref{eq:rtt_fixed} and \eqref{eq:rtt_material} is dictated by the physics. As the following examples show, it determines which time derivative appears inside the Rayleighian, and thereby the structure of the resulting evolution equations. The Navier--Stokes family of Section~\ref{sec:nsfamily} is generated by this choice.

\section{The Navier--Stokes Family}
\label{sec:nsfamily}

The first test of the framework is the family of incompressible viscous flows~--- Stokes flow, unsteady Stokes flow, and Navier--Stokes flow~--- taken in order of increasing complexity. The three members share the dissipation and the incompressibility constraint; they differ only in the \chg{kinetic energy rate} that the Rayleighian carries, and that rate is fixed by the Reynolds transport theorem: none for Stokes flow, the fixed-volume rate \eqref{eq:rtt_fixed} for unsteady Stokes flow, the material-volume rate \eqref{eq:rtt_material} for Navier--Stokes flow. Each step adds one term to the Rayleighian: the first preserves the minimum character of the principle; the second seems to reduce it to stationarity, until a change of perspective at the end of Section~\ref{sec:ns} restores it.

\subsection{Stokes flow~--- pure dissipation}
\label{sec:stokes}

The simplest member is Stokes flow, which involves only dissipation with no energy rate. It is therefore associated with the minimum family of variational principles. The Rayleighian functional consists solely of the dissipation and an incompressibility constraint:
\begin{align}
    \mathcal{R}[\boldsymbol{u}, p] = \int_V \left( \frac{1}{2}\eta\,|\nabla \boldsymbol{u}|^2 - p\,\nabla \cdot \boldsymbol{u} \right) dV\chg{,} \label{eq:stokes_L}
\end{align}
where $\boldsymbol{u}$ is the velocity field and $\eta$ the dynamic viscosity. The pressure $p$ acts as the Lagrange multiplier enforcing the incompressibility constraint $\nabla \cdot \boldsymbol{u} = 0$; as a multiplier it carries no sign restriction and is determined only up to an additive constant.

The Rayleighian $\mathscr{R} = \frac{1}{2}\eta\,|\nabla \boldsymbol{u}|^2 - p\,\nabla \cdot \boldsymbol{u}$ depends on $\boldsymbol{u}$ only through $\nabla \boldsymbol{u}$ and $\nabla \cdot \boldsymbol{u}$; both are spatial \chg{derivatives}, so integration by parts introduces a minus sign for each and the spatial Euler--Lagrange equation with respect to $\boldsymbol{u}$ reads
\begin{align}
    \chgb{-\nabla \cdot \frac{\partial \mathscr{R}}{\partial (\nabla \boldsymbol{u})} - \nabla \frac{\partial \mathscr{R}}{\partial (\nabla \cdot \boldsymbol{u})} = 0.} \label{eq:stokes_EL}
\end{align}
With $\partial \mathscr{R}/\partial (\nabla \boldsymbol{u}) = \eta\,\nabla \boldsymbol{u}$ and $\partial \mathscr{R}/\partial (\nabla \cdot \boldsymbol{u}) = -p$, this and the variation with respect to $p$ give
\begin{align}
    -\eta\,\nabla^2 \boldsymbol{u} + \nabla p = \boldsymbol{0}, \qquad \nabla \cdot \boldsymbol{u} = 0\chg{,} \label{eq:stokes}
\end{align}
the Stokes equations for creeping flow; the derivation involves only spatial operations, and no time derivatives appear. The relation to variational principles and the Rayleighian was recently discussed in \cite{beris2024}.

Helmholtz \cite{helmholtz1868} stated~--- with the first complete proof given by Korteweg \cite{korteweg1883}~--- that the Stokes solution minimizes the viscous dissipation among all incompressible velocity fields satisfying the boundary conditions.

\subsection{Unsteady Stokes flow~--- fixed control volume}
\label{sec:ustokes}

The next member adds inertia in its simplest form. Consider the kinetic energy of a fluid of constant density $\rho$ in a control volume $V$ fixed in space,
\begin{align}
    \mathcal{K} = \int_V \frac{1}{2}\rho\,|\boldsymbol{u}|^2 \, dV, \label{eq:K_fixed}
\end{align}
whose rate follows from the fixed-volume transport theorem \eqref{eq:rtt_fixed}:
\begin{align}
    \frac{d\mathcal{K}}{dt} = \int_V \rho\, \boldsymbol{u} \cdot \frac{\partial \boldsymbol{u}}{\partial t} \, dV\chg{.} \label{eq:K_rate_fixed}
\end{align}
This is the canonical TEP form \eqref{eq:F_rate}: linear in the rate $\boldsymbol{u}$, with $\rho\,\partial\boldsymbol{u}/\partial t$ in the role of the driving force. That force is the time derivative of the varied field, which an instantaneous variation of $\boldsymbol{u}$ cannot reach; it enters the Rayleighian as given data:
\begin{align}
    \mathcal{R}[\boldsymbol{u}, p] = \int_{V} \left( \rho\, \boldsymbol{u} \cdot \frac{\partial \boldsymbol{u}}{\partial t} + \frac{1}{2}\eta\,|\nabla \boldsymbol{u}|^2 - p\,\nabla \cdot \boldsymbol{u} \right) dV\chg{.} \label{eq:ustokes_L}
\end{align}
Varying $\boldsymbol{u}$ and $p$ exactly as in Section~\ref{sec:stokes} gives
\begin{align}
    \rho\,\frac{\partial \boldsymbol{u}}{\partial t} - \eta\,\nabla^2 \boldsymbol{u} + \nabla p = \boldsymbol{0}, \qquad \nabla \cdot \boldsymbol{u} = 0\chg{,} \label{eq:ustokes}
\end{align}
the unsteady Stokes equations.

\textbf{Remark on extremal character.} Nothing is frozen by fiat here: the driving force does not depend on the varied rate, so at each instant \eqref{eq:ustokes_L} is the Stokes functional \eqref{eq:stokes_L} augmented by a linear term with a given coefficient, and its minimum over divergence-free fields satisfying the boundary conditions is genuine~--- Helmholtz's theorem with the inertial force $-\rho\,\partial\boldsymbol{u}/\partial t$ as a d'Alembert body force. It is also the continuous form of the incremental minimization behind an implicit time step: with $\partial\boldsymbol{u}/\partial t \approx (\boldsymbol{u} - \boldsymbol{u}^n)/\Delta t$, the field $\boldsymbol{u}^{n+1}$ minimizes $\int_V \big[\rho\,|\boldsymbol{u}-\boldsymbol{u}^n|^2/(2\Delta t) + \tfrac{1}{2}\eta\,|\nabla\boldsymbol{u}|^2\big]\,dV$ over the same class of fields. Both regimes of Section~\ref{sec:minmax} \chgb{are described by} this one equation\chgb{, as the lid-driven cavity computation of Figure~\ref{fig:cavity} illustrates}. Under steady driving $\partial\boldsymbol{u}/\partial t \to 0$, the linear term drops out, and the minimum is Helmholtz's: the driven stationary state of Section~\ref{sec:driven}. When the driving stops, the flow in a closed container with no-slip walls decays with $\dot{\mathcal{K}} + \mathcal{P} = 0$: the stored kinetic energy is released, at given driving force $\rho\,\partial\boldsymbol{u}/\partial t$, through the rates of maximal dissipation~--- the relaxation regime of Section~\ref{sec:relaxation}.

\subsection{Navier--Stokes flow~--- material control volume}
\label{sec:ns}

The unsteady Stokes equation misses the convective nonlinearity, which no fixed control volume can supply: the fluid moves, and the kinetic energy it carries is that of a material volume $V(t)$:
\begin{align}
    \mathcal{K} = \int_{V(t)} \frac{1}{2}\rho\,|\boldsymbol{u}|^2 \, dV\chg{.} \label{eq:K_material}
\end{align}
By the Reynolds transport theorem for a material volume \eqref{eq:rtt_material}, with $\nabla \cdot \boldsymbol{u} = 0$ (incompressibility):
\begin{align}
    \frac{d\mathcal{K}}{dt} = \int_{V(t)} \rho\, \boldsymbol{u} \cdot \frac{D\boldsymbol{u}}{Dt} \, dV\chg{,} \label{eq:K_rate}
\end{align}
where $D\boldsymbol{u}/Dt = \partial\boldsymbol{u}/\partial t + (\boldsymbol{u} \cdot \nabla)\boldsymbol{u}$ is the material derivative of the velocity. The Rayleighian functional combines this rate with the dissipation:
\begin{align}
    \mathcal{R}[\boldsymbol{u}, p] = \int_{V(t)} \left( \boldsymbol{u} \cdot \textcolor{freezepink}{\Big( \rho\, \frac{D\boldsymbol{u}}{Dt} \Big)_{\!\text{freeze}}} + \frac{1}{2}\eta\,|\nabla \boldsymbol{u}|^2 - p\,\nabla \cdot \boldsymbol{u} \right) dV\chg{.} \label{eq:ns_L}
\end{align}

\chg{\textbf{Key point.}} The material derivative $D\boldsymbol{u}/Dt$ in \eqref{eq:ns_L} is \textit{not varied}; it enters as given data~--- the fluid's ``acceleration''~--- much as external forces do in Hamilton's principle, and as $\partial\boldsymbol{u}/\partial t$ did in Section~\ref{sec:ustokes}. Here the TEP loses some of its elegance. For inertialess systems the driving force is a state function, and for unsteady Stokes flow it is the time derivative of the varied field, which the variation cannot reach. In the Eulerian description of Navier--Stokes flow it contains the varied rate itself, through the convective term $(\boldsymbol{u}\cdot\nabla)\boldsymbol{u}$, and has to be held fixed by hand; the end of this section shows that this is unnecessary in material coordinates. Section~\ref{sec:lagrangian} shows that the same rule, applied to the total energy of Lagrangian mechanics, yields Lagrange's equations with dissipative forces. Varying only the explicit occurrences of $\boldsymbol{u}$:
\begin{align}
    \rho\,\frac{D\boldsymbol{u}}{Dt} - \eta\,\nabla^2 \boldsymbol{u} + \nabla p = \boldsymbol{0}\chg{.} \label{eq:ns}
\end{align}
This is the Navier--Stokes equation. Together with $\nabla \cdot \boldsymbol{u} = 0$, it governs incompressible viscous flow.

\textbf{Remark on extremal character.} In a variational principle every term of the resulting equation is the derivative of a term of the functional. In \eqref{eq:ns} the viscous term is the derivative of the dissipation \chg{potential} $\tfrac{1}{2}\eta\,|\nabla\boldsymbol{u}|^2$ with respect to $\boldsymbol{u}$, and the pressure gradient is that of the constraint term. A term of an equation is called \textit{variational} if it is the derivative of some functional, in the same way as a conservative force is the gradient of a potential; a term for which no such functional exists is \textit{\chg{nonvariational}}. The convective term $\rho(\boldsymbol{u}\cdot\nabla)\boldsymbol{u}$ is of the second kind, a classical result \cite{finlayson1972}. The natural candidate for its functional is the difference between the two energy rates \eqref{eq:K_rate} and \eqref{eq:K_rate_fixed}, $\int_V \rho\,\boldsymbol{u}\cdot(\boldsymbol{u}\cdot\nabla)\boldsymbol{u}\, dV$. Its integrand is a pure divergence, $\nabla\cdot(\tfrac{1}{2}\rho|\boldsymbol{u}|^2\boldsymbol{u})$, so that the integral equals the boundary term $\oint_{\partial V} \tfrac{1}{2}\rho|\boldsymbol{u}|^2\, \boldsymbol{u}\cdot\boldsymbol{n}\, dS$, which vanishes for $\boldsymbol{u}\cdot\boldsymbol{n} = 0$ on $\partial V$. The candidate is therefore zero for every admissible field, and its derivative is zero as well. The same holds for any other functional: since the convective term is quadratic in $\boldsymbol{u}$, a functional $\Phi$ with this term as its derivative would satisfy $\Phi[\boldsymbol{u}] - \Phi[\boldsymbol{0}] = \tfrac{1}{3}\int_V \rho\,\boldsymbol{u}\cdot(\boldsymbol{u}\cdot\nabla)\boldsymbol{u}\, dV = 0$ and would be constant. Physically, the convective term does no net work in a closed container; it only redistributes kinetic energy within the fluid. It can therefore enter \eqref{eq:ns} in one way only, as part of the given data: it is frozen inside $D\boldsymbol{u}/Dt$, although it depends explicitly on the varied field. In this sense the Eulerian inertial driving force is artificial: as long as $\boldsymbol{u}$ is varied at fixed $\partial\boldsymbol{u}/\partial t$, the frozen force contains the varied field, the class of competing fields is defined relative to the solution, and no minimum statement is available; the stationarity condition alone generates \eqref{eq:ns}. Section~\ref{sec:lagrangian} \chgb{discusses} the same obstacle in mechanics. The obstacle belongs to the Eulerian description and disappears under a change of perspective.

\textbf{Change of perspective: material coordinates.} Describe the motion by the map $\boldsymbol{x} = \boldsymbol{\chi}(\boldsymbol{X}, t)$ from the reference configuration $V_0$ to the current one, so that the state is $\boldsymbol{\chi}$ and the rate is the material velocity $\dot{\boldsymbol{\chi}}$. The Eulerian velocity is the velocity of whichever particle occupies $\boldsymbol{x}$, and the chain rule identifies the particle acceleration with the material derivative:
\begin{align}
    \boldsymbol{u} = \dot{\boldsymbol{\chi}} \circ \boldsymbol{\chi}^{-1}, \qquad
    \ddot{\boldsymbol{\chi}} \circ \boldsymbol{\chi}^{-1} = \frac{\partial \boldsymbol{u}}{\partial t} + (\boldsymbol{u} \cdot \nabla)\boldsymbol{u} = \frac{D\boldsymbol{u}}{Dt}\chg{.} \label{eq:material_map}
\end{align}
($\circ$ denotes composition in the space variable at fixed $t$). With $\rho_0\, dV_0 = \rho\, dV$ the kinetic energy \eqref{eq:K_material} and its rate \eqref{eq:K_rate} read
\begin{align}
    \mathcal{K} = \int_{V_0} \tfrac{1}{2}\rho_0\, |\dot{\boldsymbol{\chi}}|^2 \, dV_0, \qquad
    \frac{d\mathcal{K}}{dt} = \int_{V_0} \rho_0\, \dot{\boldsymbol{\chi}} \cdot \ddot{\boldsymbol{\chi}} \, dV_0, \label{eq:K_material_coords}
\end{align}
and the transport theorem has become trivial\chgb{, since} $V_0$ and $\rho_0(\boldsymbol{X})$ do not move. The coefficient of the rate is now $\rho_0\, \ddot{\boldsymbol{\chi}}$, the second time derivative of the state, which an instantaneous variation of $\dot{\boldsymbol{\chi}}$ cannot reach. This is the situation of Section~\ref{sec:ustokes}, and nothing has to be held fixed by hand. At frozen $\boldsymbol{\chi}$ the map $\dot{\boldsymbol{\chi}} \mapsto \boldsymbol{u}$ is linear, the constraint $\det \nabla_{\!X}\boldsymbol{\chi} = 1$ has the rate form $\nabla \cdot \boldsymbol{u} = 0$, and the dissipation is quadratic in $\dot{\boldsymbol{\chi}}$; written in the current configuration, the material Rayleighian is therefore exactly \eqref{eq:ns_L}, with $D\boldsymbol{u}/Dt$ supplied by \eqref{eq:material_map}, and its variation is \eqref{eq:ns}. The convective term arises only when the particle acceleration is expressed through the Eulerian derivatives. \chgb{This yields one of the main results of this work:}
\begin{quote}
    \chgb{In material coordinates,} at given particle acceleration the functional \eqref{eq:ns_L} is strictly convex in $\boldsymbol{u}$ over divergence-free fields satisfying the boundary conditions, and the Navier--Stokes solution is its genuine minimum, on the same footing as the unsteady Stokes minimum at given $\partial\boldsymbol{u}/\partial t$.
\end{quote}
For a closed container ($\boldsymbol{u} \cdot \boldsymbol{n} = 0$) the two descriptions even agree on the value of the energy rate, since the rates \eqref{eq:K_rate} and \eqref{eq:K_rate_fixed} differ by $\oint_{\partial V} \tfrac{1}{2}\rho|\boldsymbol{u}|^2\, \boldsymbol{u}\cdot\boldsymbol{n}\, dS = 0$; the two Rayleighians differ only in which part of the particle acceleration is treated as data, the local rate $\partial\boldsymbol{u}/\partial t$ or all of $D\boldsymbol{u}/Dt$. The two inertial members of the family therefore differ in the acceleration that is held fixed: the local one gives the unsteady Stokes equation, the particle acceleration gives the Navier--Stokes equation.

\section{Dissipative Lagrangian Mechanics}
\label{sec:lagrangian}

The final example connects the TEP to classical mechanics. The TEP demands a free energy that is a \textit{state function} whose rate is linear in the varied rates, as in \eqref{eq:F_rate}. For a system with inertia this seems to fail: the kinetic energy depends on the velocity\chg{, that is,} on the rate itself\chg{,} so the driving force in $\dot{\mathcal{F}}$ is no longer a state function: it contains the acceleration of the varied variable, as $\partial\boldsymbol{u}/\partial t$ did in Section~\ref{sec:ustokes}, and for a general kinetic metric the velocity itself, as $D\boldsymbol{u}/Dt$ did in Section~\ref{sec:ns}. The usual response is to restrict the Rayleighian extremum to the overdamped regime, where the kinetic energy is negligible \cite{doi2011, beris2024}; Beris and Edwards show explicitly that their extremum principle for the damped oscillator recovers only the inertialess equation \cite{beris2024}. \chgb{In the present work, the} restriction is lifted, as \chgb{for the fluids of} Section~\ref{sec:nsfamily}, by the choice of free energy. For a mechanical system at macroscopic scale\chg{,} as opposed to a microscopic one such as the atoms of an ideal gas, whose free energy carries an entropic part\chg{,}\footnote{Boltzmann did bridge the two: statistical mechanics derives temperature and entropy from \chgb{first principles}, starting from the Hamiltonian, and predicts the thermodynamic properties of an ideal gas from its classical Hamiltonian alone. The price is ergodicity: temperature can be assigned to a classical ensemble only if the system is very large \chg{(}many degrees of freedom\chg{)} and chaotic in a definite sense. For a \chg{mechanical system with few degrees of freedom, such as a pendulum}, temperature and entropy are therefore not applicable; see Chapter~2 of \cite{berdichevsky2009}.} the free energy is the total energy $\chg{\mathscr{E}}$, which we define in terms of the Lagrangian to simplify the derivations:
\begin{align}
    \chg{\mathscr{E}}(q, \dot{q}) := \sum_i \dot{q}_i\, \frac{\partial \mathscr{L}}{\partial \dot{q}_i} - \mathscr{L}(q, \dot{q}), \label{eq:legendre}
\end{align}
which is $K + \chg{\Pi}$ for $\mathscr{L} = K - \chg{\Pi}$ with $K$ quadratic in the velocities \chg{and $\Pi$ the potential energy}. The dissipation is \chgb{assumed to be} that of linear dampers with constants $\chg{c_i}$: Rayleigh's dissipation function $\sum_i \tfrac{1}{2} \chg{c_i} \dot{q}_i^2$ \cite{rayleigh1873} plays the part of $\tfrac{1}{2}\mathcal{P}$ in \eqref{eq:tep_general}. The total rate of $\chg{\mathscr{E}}$ along a trajectory is
\begin{align}
    \frac{d\chg{\mathscr{E}}}{dt}
    = \sum_i \Big( \frac{d}{dt}\frac{\partial \mathscr{L}}{\partial \dot{q}_i} - \frac{\partial \mathscr{L}}{\partial q_i} \Big) \dot{q}_i
    = \sum_i X_i\, \dot{q}_i \label{eq:H_rate_legendre}
\end{align}
\textit{exactly}, for any Lagrangian \chg{without explicit time dependence} and any number of degrees of freedom: the acceleration terms $\sum_i (\partial \mathscr{L}/\partial \dot{q}_i)\, \ddot{q}_i$ arising from the two \chg{terms} of \eqref{eq:legendre} cancel, and what remains is the Euler--Lagrange expression times the velocity~--- the canonical TEP form \eqref{eq:F_rate}, linear in the varied rates. The acceleration is still present inside the coefficients $X_i$, and for a general Lagrangian so is the velocity. The $X_i$ are therefore held fixed during the variation~--- the Euler--Lagrange expression enters as trajectory data, exactly as $\rho\, D\boldsymbol{u}/Dt$ does in Section~\ref{sec:ns}. This is less natural than in the inertialess case, where the $X_i$ are state functions and nothing is frozen by hand. An instantaneous rate principle requires it whenever the energy of the system depends on the rates. With this rule the variation proceeds as in every other example:
\begin{align}
    \delta_{\dot{q}} \bigg[ \sum_i \textcolor{freezepink}{\Big( \frac{d}{dt}\frac{\partial \mathscr{L}}{\partial \dot{q}_i} - \frac{\partial \mathscr{L}}{\partial q_i} \Big)_{\!\text{freeze}}}\, \dot{q}_i + \sum_i \tfrac{1}{2} \chg{c_i} \dot{q}_i^2 \bigg] = 0
    \quad \Longrightarrow \quad
    \frac{d}{dt} \frac{\partial \mathscr{L}}{\partial \dot{q}_i} - \frac{\partial \mathscr{L}}{\partial q_i}
    = -\chg{c_i}\, \dot{q}_i \quad \forall\, i\chg{.} \label{eq:lagrange_rayleigh}
\end{align}
These are Lagrange's equations with \chg{the generalized friction forces $-c_i\dot{q}_i$ obtained from} Rayleigh's dissipation function~--- the structure \textit{postulated} by Rayleigh in 1873 \cite{rayleigh1873}, here \textit{derived} from the TEP.

\textbf{When the frozen force is exact: the connection to the fluid examples.} The frozen-force rule can be checked by direct calculation. For a single degree of freedom (several add only index bookkeeping and leave the conclusion unchanged), writing out the time derivative in \eqref{eq:H_rate_legendre} gives
\begin{align}
    \chg{\dot{\mathscr{E}}} = \dot{q} \left( \frac{\partial^2 \mathscr{L}}{\partial \dot{q}^2}\, \ddot{q}
        + \frac{\partial^2 \mathscr{L}}{\partial q\, \partial \dot{q}}\, \dot{q}
        - \frac{\partial \mathscr{L}}{\partial q} \right), \label{eq:X_explicit}
\end{align}
an explicit function of the state $q$, the varied rate $\dot{q}$\chg{,} and the acceleration $\ddot{q}$. An instantaneous variation of $\dot{q}$ leaves $q$ and $\ddot{q}$ untouched~--- the acceleration \chgb{behaves} exactly as $\partial\boldsymbol{u}/\partial t$ \chgb{did} in Section~\ref{sec:ustokes}~--- so the \chg{full} derivative of $\chg{\dot{\mathscr{E}}}$ with respect to the varied rate is
\begin{align}
    \frac{\partial \chg{\dot{\mathscr{E}}}}{\partial \dot{q}}\bigg|_{q,\ddot{q}}
    = \frac{d}{dt}\frac{\partial \mathscr{L}}{\partial \dot{q}} - \frac{\partial \mathscr{L}}{\partial q}
    + \dot{q} \left( \frac{\partial^3 \mathscr{L}}{\partial \dot{q}^2\, \partial q}\, \dot{q}
        + \frac{\partial^3 \mathscr{L}}{\partial \dot{q}^3}\, \ddot{q} \right). \label{eq:honest_variation}
\end{align}
The frozen-force rule behind \eqref{eq:lagrange_rayleigh} consists in discarding the last term of \eqref{eq:honest_variation}. It is exact when that term vanishes, that is, when the driving force does not depend on the varied rate; nothing is then frozen by fiat. For $\mathscr{L} = K - \chg{\Pi}$ with $\chg{\Pi = \Pi(q)}$ this happens if and only if the kinetic energy depends on the velocity alone, $K = K(\dot{q})$, quadratic with a constant coefficient: then both third derivatives vanish identically, $X = m\ddot{q} + \chg{\Pi'(q)}$ with constant $m$, and the \chg{full} variation $\delta_{\dot{q}}[\chg{\dot{\mathscr{E}}} + \tfrac{1}{2} \chg{c} \dot{q}^2] = 0$ at fixed $(q, \ddot{q})$ returns \eqref{eq:lagrange_rayleigh} unaided.

This is the mechanical counterpart of the two inertial members of the Navier--Stokes family. A kinetic energy that depends on the velocities alone is the fixed-volume kinetic energy \eqref{eq:K_fixed}, and equally the material-volume kinetic energy in material coordinates \eqref{eq:K_material_coords}: the coefficient of the rate does not depend on the rate, the variation is exact, and the result is the unsteady Stokes equation at given $\partial\boldsymbol{u}/\partial t$ or the Navier--Stokes equation at given particle acceleration, each with its genuine minimum (Sections~\ref{sec:ustokes} and~\ref{sec:ns}). The Eulerian form of the material-volume kinetic energy plays the part of a configuration-dependent mass matrix. Its coefficient is the constant $\rho$, but its driving force, the particle acceleration expressed through the Eulerian derivatives, acquires the convective term, quadratic in $\boldsymbol{u}$, which must be frozen (\chgb{see} Section~\ref{sec:ns}), \chg{just} as the term $\tfrac{1}{2} m'(q)\dot{q}^2$ must \chg{be frozen when the mass $m(q)$ depends on the position}. The distinction between the Eulerian and the material description of Section~\ref{sec:ns}, namely whether the coefficient of the rate depends on the rate, corresponds in mechanics to the distinction between $K = K(\dot{q})$ and $K = K(q, \dot{q})$ in the chosen coordinates. For $K = K(\dot{q})$ the TEP is the same \chg{self-contained}, exact principle as for \chg{unsteady} Stokes flow: the minimum of the Rayleighian at fixed $(q, \ddot{q})$ is genuine, and the damped system operates in the constrained-maximum regime of Section~\ref{sec:relaxation}.

A closely related \chgb{solution} was proposed independently by Hackl, Svoboda\chg{,} and Fischer \cite{hackl2024coupling}, who extend the Lagrangian by a dissipation-force term and couple a Modified Hamilton's Principle with the TEP through the Ziegler form. Both derivations recover Lagrange's equations with Rayleigh dissipation, and both confront the double role of $\dot{q}$ as state and as rate; here the rate of the total energy is the Euler--Lagrange expression times the velocity~--- already the canonical TEP form~--- and the rule used for $\rho\, D\boldsymbol{u}/Dt$ in Section~\ref{sec:ns}, driving forces frozen during the variation, does the rest, so no extended Lagrangian or auxiliary principle is required.

\textbf{Observation: spontaneous synchronization.} The constant-mass minimum has a visible consequence. Coupled self-sustained oscillators~--- Huygens' pendulum clocks on a common beam \cite{bennett2002}, metronomes on a common support \cite{pantaleone2002}, phase-locked oscillator networks \cite{pikovsky2001}~--- run in the driven regime of Section~\ref{sec:driven}: the escapements supply on average what the damping dissipates, so $\chg{\dot{\mathscr{E}}} \approx 0$ over a cycle. Averaged over the fast oscillation, the slow drift of the phase difference $\Delta\varphi$ \chg{can be read as a} gradient descent of the averaged coupling dissipation. For two oscillators \chg{of amplitude $A$ and frequency $\omega_0$} coupled by a dashpot $c$, $\langle \mathcal{P}_c \rangle = c A^2 \omega_0^2 (1 - \cos\Delta\varphi)$ is minimized in phase; for Huygens' clocks the damped beam couples to the sum of the momenta, $\langle \mathcal{P}_{\mathrm{beam}} \rangle \propto 1 + \cos\Delta\varphi$ is minimized in anti-phase, and the beam stands still \cite{bennett2002}. In both cases the synchronized pattern extinguishes the collective mode that carries the coupling dissipation. Synchronization is the constrained minimum of Section~\ref{sec:driven} for coupled oscillators, and the selected pattern is determined by the shared mode that the coupling damps, whatever its frequency. \chgb{A speculative extension to composite systems is suggested by the lateral sway of London's Millennium Bridge on its opening day \cite{strogatz2005}. The bridge, described by linear continuum mechanics, has a constant mass matrix, the case $K = K(\dot{q})$ above, so the extremal statements apply to it as they do to unsteady Stokes flow. The pedestrians are highly nonlinear thermodynamic subsystems, but each of them solves an optimal control problem, moving with as little effort, that is, as little dissipation, as possible. Assembled from such subsystems and the bridge, the whole system would then tend toward a dynamical minimum: a nonequilibrium stationary state in which its mass and energy stay constant and the dissipation is minimized, as for Stokes flow, with the crowd falling into step as the visible sign.}

\section{Discussion}
\label{sec:discussion}

\subsection{On minima and maxima of the TEP}
\label{sec:minmax}

Whether the TEP is a principle of minimum or of maximum dissipation has long been debated \cite{martyushev2006, struchtrup1998, polettini2013}; for multicomponent diffusion, a recent \chgb{paper} by \chg{Svoboda, Hackl, and Fischer} \cite{svoboda2023note} showed \chgb{the maximum and the minimum principle} to be different extremization tasks\chg{, in which the rates are varied at given forces or the fields under fixed boundary values; for the systems considered here they correspond to} the two regimes below. The general statement rests on one observation:
\begin{quote}
    The extremal character is decided by the \textit{regime}, that is, by the kind of thermodynamic system at hand: in a closed system relaxing at frozen state the dissipation is \textit{maximized} over the balance-consistent rates; at the driven stationary state of an open system it is \textit{minimized} over the fields compatible with the driving.
\end{quote}

\subsubsection{Relaxation at frozen state: constrained maximum}
\label{sec:relaxation}

Consider a closed system releasing free energy at frozen state (given forces $X_i$), as in Section~\ref{sec:derivation} (Figure~\ref{fig:systems}c). The rates admitted to the comparison are all those that satisfy the balance $\dot{\mathcal{F}} + \mathcal{P} = 0$. They are trial rates: each of them accounts for the free energy released per unit time, and only the actual rates also obey the Onsager law \eqref{eq:onsager_law}. The balance defines an ellipsoid through the origin, and the dissipation attains its nontrivial constrained extremum, a \textit{maximum}, at the actual rates. This is Ziegler's principle of maximum dissipation \cite{ziegler1963}, which at uniform constant temperature coincides with maximum entropy production \cite{martyushev2006} since $\mathcal{P} = \chg{T\, d_i S/dt}$. Note the logic: the energy balance is \textit{assumed}, and the extremum condition \chg{\textit{yields}} the Onsager law, which is the constitutive law of the system. Unsteady Stokes and Navier--Stokes flow decaying after the driving has stopped operate in this regime (Sections~\ref{sec:ustokes} and~\ref{sec:ns}, the latter in material coordinates), as does damped mechanics with a constant mass matrix; with a configuration-dependent kinetic metric no extremal statement is available in the given coordinates (Section~\ref{sec:lagrangian}).

\subsubsection{Driven \texorpdfstring{\chg{stationary}}{stationary} states: constrained minimum}
\label{sec:driven}

\begin{figure}[htbp]
\centering
\resizebox{\textwidth}{!}{%
\begin{tikzpicture}
% ===== (a) open system, \chg{nonstationary} =====
\begin{scope}[xshift=0cm]
\draw[thick] (-3.0, -2.75) rectangle (3.0, 2.75);
\node at (0, 2.3) {\small\bfseries (a) Open, \chg{nonstationary}};
\node at (0, 1.5) {$\chg{\dot{\mathcal{U}}} \neq 0, \qquad \dfrac{d_i S}{dt} \geq 0$};
\node at (0, 0.7) {$\dot{\mathcal{F}} + \mathcal{P} = \dot{W}_{\text{ext}}, \quad \dot{\mathcal{F}} \neq 0$};
\node at (0, -0.05) {\small $\mathcal{R}$ minimized at given acceleration;};
\node at (0, -0.45) {\small \chg{driven through prescribed rates:}};
\node at (0, -1.1) {$\to$ \chg{\textbf{no extremum} of $\mathcal{P}$}};
\node at (0, -1.75) {\small (TEP generates the equations;};
\node at (0, -2.15) {\small energy is being stored or released)};
\draw[->, very thick, red] (-1.7, 3.75) -- (-1.7, 2.8);
\node[red] at (-0.85, 3.3) {\small $\dot{W}_{\text{ext}}$};
\draw[->, very thick, blue] (1.7, -2.8) -- (1.7, -3.75);
\node[blue] at (0.35, -3.3) {\small $d_e S/dt < 0$};
\end{scope}
% ===== (b) open system, stationary state =====
\begin{scope}[xshift=6.8cm]
\draw[thick] (-3.0, -2.75) rectangle (3.0, 2.75);
\node at (0, 2.3) {\small\bfseries (b) Open, stationary};
\node at (0, 1.5) {$\chg{\dot{\mathcal{U}}} = 0, \qquad \dfrac{d_i S}{dt} = -\dfrac{d_e S}{dt}$};
\node at (0, 0.7) {$\chg{dS/dt} = 0, \quad \mathcal{P} = \dot{W}_{\text{ext}}$};
\node at (0, -0.05) {\small $\mathcal{R} \to \tfrac{1}{2}\mathcal{P}$: \ $\mathcal{P}$ extremized over};
\node at (0, -0.45) {\small driving-constrained fields};
\node at (0, -1.1) {$\to$ \textbf{minimum} \small($\mathcal{P}$ convex)};
\node at (0, -1.75) {\small (Prigogine: minimum entropy};
\node at (0, -2.15) {\small production, stationary state)};
\draw[->, very thick, red] (-1.7, 3.75) -- (-1.7, 2.8);
\node[red] at (-0.85, 3.3) {\small $\dot{W}_{\text{ext}}$};
\draw[->, very thick, blue] (1.7, -2.8) -- (1.7, -3.75);
\node[blue] at (0.35, -3.3) {\small $d_e S/dt < 0$};
\end{scope}
% ===== (c) closed system, relaxing =====
\begin{scope}[xshift=13.6cm]
\draw[thick] (-3.0, -2.75) rectangle (3.0, 2.75);
\node at (0, 2.3) {\small\bfseries (c) Closed, relaxing};
\node at (0, 1.5) {$\chg{\dot{W}_{\text{ext}} = 0}, \qquad \dfrac{d_i S}{dt} \geq 0$};
\node at (0, 0.7) {$\dot{\mathcal{F}} + \mathcal{P} = 0, \quad \dot{\mathcal{F}} < 0$};
\node at (0, -0.05) {\small $\mathcal{P}$ extremized over};
\node at (0, -0.45) {\small balance-consistent rates};
\node at (0, -1.1) {$\to$ \textbf{maximum}};
\node at (0, -1.75) {\small (Onsager law derived;};
\node at (0, -2.15) {\small \chg{nonstationary} state)};
\draw[->, very thick, blue, chg] (1.7, -2.8) -- (1.7, -3.75);
\node[blue, chg] at (0.1, -3.3) {\small $d_e S/dt = \dot{Q}/T$};
\end{scope}
\end{tikzpicture}}
\caption{The three regimes of a dissipative system in Prigogine's notation ($dS = d_i S + d_e S$, second law $d_i S/dt \geq 0$, dissipation $\mathcal{P} = T\, d_i S/dt$), \chg{labeled} like the three phases of the cavity computation of Figure~\ref{fig:cavity}. (a)~An open system in a \chg{nonstationary} state: energy is supplied and partly stored or released, $\dot{\mathcal{F}} + \mathcal{P} = \dot{W}_{\text{ext}}$ with $\dot{\mathcal{F}} \neq 0$; the Rayleighian is minimized at the given acceleration and generates the evolution equations, but no extremal statement about the dissipation follows \chg{when the system is driven through prescribed rates, as the cavity is; when it is driven through prescribed forces or potentials, the maximum of (c) survives (Section~\ref{sec:driven})}. (b)~An open system in a nonequilibrium stationary state: the stored energy is constant ($\chg{\dot{\mathcal{U}}} = 0$), the produced entropy is exported ($d_i S = -d_e S$, so $dS = 0$), and the supplied power is dissipated, $\mathcal{P} = \dot{W}_{\text{ext}}$; the Rayleighian functional reduces to $\tfrac{1}{2}\mathcal{P}$ and the actual fields \textit{minimize} the dissipation over the driving-constrained class~--- Prigogine's minimum entropy production (Section~\ref{sec:driven}). (c)~A closed system \chg{at constant temperature, supplied with no work ($\dot{W}_{\text{ext}} = 0$) and exchanging heat $\dot{Q}$ with a bath,} relaxing through \chg{nonstationary} states: the free energy pays for the dissipation, $\dot{\mathcal{F}} + \mathcal{P} = 0$, and among all rates consistent with this balance the actual ones \textit{maximize} the dissipation~--- maximum entropy production at the \chg{nonstationary} state (Section~\ref{sec:relaxation}). The extremal character is decided by the regime; in all three $\mathcal{R} = \dot{\mathcal{F}} + \tfrac{1}{2}\mathcal{P}$ is minimized, and only in (b) and (c) does this translate into an extremal statement about $\mathcal{P}$ itself\chg{, unless (a) is driven through prescribed forces}.}
\label{fig:systems}
\end{figure}
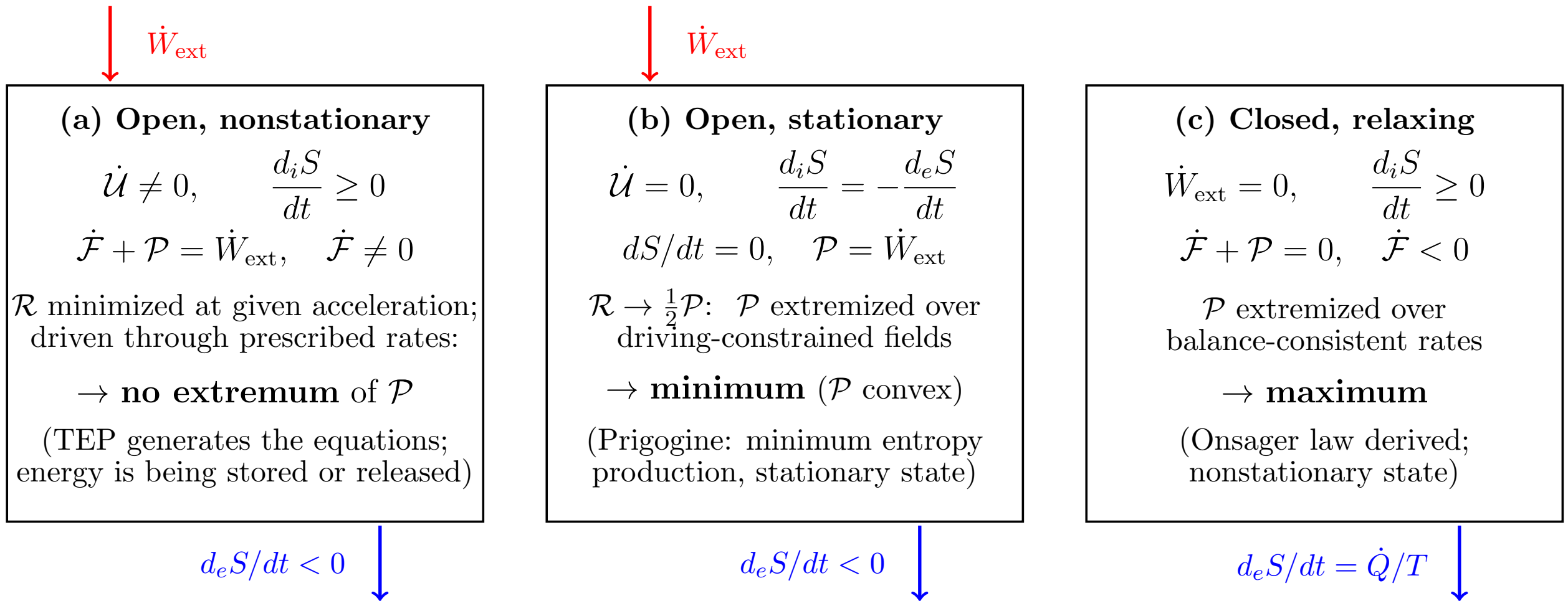

Now consider a system driven by external power input and exporting entropy to its surroundings (Figure~\ref{fig:systems}b). The balance generalizes to include the external power $\dot{W}_{\text{ext}}$:
\begin{align}
    \dot{\mathcal{F}} + \mathcal{P} = \dot{W}_{\text{ext}}\chg{.} \label{eq:balance_driven}
\end{align}
\chg{If the power is supplied through prescribed forces or potentials, as in the diffusion problem of Svoboda et al.\ \cite{svoboda2023note}, it is linear in the rates, $\dot{W}_{\text{ext}} = \sum_i Y_i \dot{q}_i$ with frozen $Y_i$. The balance \eqref{eq:balance_driven} is then again an ellipsoid through the origin, with $X_i$ replaced by $X_i - Y_i$, and the maximum statement of Section~\ref{sec:relaxation} holds in the open system as well. \chgb{If instead the rates are prescribed on the boundary, such as the lid velocity in the cavity computation of Figure~\ref{fig:cavity}, the external power depends on boundary forces that are not known in advance; it is then not a given linear function of the rates, and the balance yields no maximum statement.}} In a \chg{stationary} state $\dot{\mathcal{F}} = 0$ and the supplied power is entirely dissipated, $\mathcal{P} = \dot{W}_{\text{ext}}$. \chg{For driving through prescribed rates, the} balance then no longer constrains the competing rates in any useful way. For Stokes flow no free energy is released in the bulk, so that the linear term of the Rayleighian vanishes, the ellipsoid of Section~\ref{sec:derivation} degenerates to the single point $\dot{q} = 0$, and no maximum statement exists. The competing fields are fixed by the \textit{driving}: all those satisfying the prescribed boundary conditions and kinematic constraints, with their dissipation evaluated using the constitutive law. On this class the Rayleighian functional reduces to $\tfrac{1}{2}\mathcal{P}$ and the physical solution is the constrained \textit{minimum}~--- Helmholtz's theorem (Section~\ref{sec:stokes}). The logic is reversed: the constitutive law is \textit{assumed}, the competing fields violate the momentum balance, and the extremum condition \chg{\textit{yields}} it. Explicitly: \textit{for an open system driven through prescribed boundary values of the rates, in which the driving forces vanish at the stationary state, the Rayleighian minimum principle degenerates to a pure minimum of the dissipation~--- at constant temperature, of the entropy production, since $\mathcal{P} = \chg{T\, d_i S/dt}$. Prigogine's minimum entropy production theorem, its counterpart for constant coefficients, is thus contained in the Rayleighian functional as this limiting case.} \chg{Svoboda et al.\ \cite{svoboda2023note} state the minimum for diffusion under the dual condition: the chemical potentials are prescribed on the surface and varied inside, so that the dissipation is minimized in its force representation.}

The two statements are dual: each answers the same question in its own regime. The unifying object is the Rayleighian functional $\mathcal{R} = \dot{\mathcal{F}} + \tfrac{1}{2}\mathcal{P}$, which is convex and minimized in \textit{every} regime: during relaxation its minimum, expressed through $\mathcal{P}$ alone on the balance surface, becomes the constrained maximum of Section~\ref{sec:relaxation}; under steady driving the energy rate drops out and the minimum of $\mathcal{R}$ becomes the minimum of $\mathcal{P}$ itself. Equilibrium thermodynamics has the same structure: one and the same equilibrium state maximizes the entropy at fixed energy and minimizes the free energy at fixed temperature, and which function is extremal depends on what is controlled \cite{kondepudi2014modern}. The two regimes, and the \chg{nonstationary rate-driven} state that belongs to neither, are summarized in Figure~\ref{fig:systems}.

\textbf{\chgb{The cavity revisited.}} \chgb{The three phases of the cavity computation of Figure~\ref{fig:cavity} are the regimes of Figure~\ref{fig:systems}. The computation solves the} unsteady Stokes \chgb{equation} \eqref{eq:ustokes} in a square cavity of side $L$, driven by a lid moving with velocity $U\,16\chg{(x/L)^2(1-x/L)^2}$ (regularized so that the dissipation stays finite at the corners), discretized with Taylor--Hood elements and advanced in time by the backward Euler method, that is, by the incremental minimization of Section~\ref{sec:ustokes}. \chgb{Phase (a) is} driven but not stationary, $\dot{\mathcal{K}} + \mathcal{P} = \dot{W}_{\mathrm{lid}}$ with all three terms different from zero, and\chg{, since the lid prescribes rates,} no extremal statement about the dissipation is available; the TEP only generates the equation. The computation shows the approach to the stationary state: the dissipation decreases monotonically towards Helmholtz's minimum $\mathcal{P}_S$, the excess $\mathcal{P} - \mathcal{P}_S$ being the dissipation of the decaying difference field, which is orthogonal to the Stokes solution in the inner product of the dissipation. \chg{That the dissipation decreases to its stationary minimum was anticipated in \cite{struchtrup1998, svoboda2023note}.} Phase (b) is the driven stationary state of Section~\ref{sec:driven}: $\dot{\mathcal{K}} = 0$ and $\mathcal{P} = \dot{W}_{\mathrm{lid}} = \mathcal{P}_S$, the constrained minimum. \chgb{In phase (c) the container is} closed, $\dot{\mathcal{K}} + \mathcal{P} = 0$, and the stored kinetic energy is released through the rates of maximal dissipation, the constrained maximum of Section~\ref{sec:relaxation}, until only the slowest Stokes mode remains; the kinetic energy \chgb{present when the lid stops} is dissipated entirely, and the discrete energy identity, including the numerical dissipation of the backward Euler step, \chgb{is satisfied to machine precision} at every time step.

\subsection{Linearity, gradient-flow structure, and the role of equilibrium}
\label{sec:linearity}

The quadratic dissipation \eqref{eq:entropy_prod} places the framework in the linear regime of nonequilibrium thermodynamics. The linearity concerns the kinetics \textit{at frozen state}: the Onsager law \eqref{eq:onsager_law} is linear in the forces, while the forces themselves, $X_i = \partial\mathcal{F}/\partial q_i$, are in general nonlinear functions of the state, and the coefficients $L_{ij}$ may depend on it as well. As an evolution equation, $\dot{q}_i = -\sum_j L_{ij}(q)\,\partial\mathcal{F}/\partial q_j$ is therefore nonlinear in general. For an inertialess system it is the gradient flow of the free energy in the metric defined by the dissipation, with $\dot{\mathcal{F}} = -\mathcal{P} \leq 0$ along every solution of a closed system, and the minimization of the Rayleighian at each instant is the time-continuous form of the minimizing-movement scheme by which such flows are constructed \cite{jordan1998, otto2001}. With inertia the driving force contains the acceleration and the evolution is no longer a gradient flow of the energy. Unsteady Stokes flow is a gradient flow with the roles exchanged: it descends the dissipation potential $\tfrac{1}{2}\mathcal{P}$ in the metric of the kinetic energy, and the implicit time step of Section~\ref{sec:ustokes} is its minimizing movement. In all these cases the kinetics is linear\chg{, while the dynamics may be} nonlinear: the potential of Section~\ref{sec:lagrangian} is arbitrary, and the Navier--Stokes equation is nonlinear through its inertial force.

The quadratic form is, moreover, more restrictive than necessary. Let the dissipation be differentiable, strictly convex, positive for nonzero rates\chg{,} and positively homogeneous of degree $k > 1$ in the rates, $\mathcal{P}(\alpha \dot{q}) = \alpha^k\, \mathcal{P}(\dot{q})$ for $\alpha > 0$, with coefficients that may again depend on the state. The constrained stationarity \eqref{eq:tep_constrained} and Euler's theorem on homogeneous functions read
\begin{align}
    (1+\lambda)\, \frac{\partial\mathcal{P}}{\partial\dot{q}_i} + \lambda X_i = 0, \qquad \sum_i \dot{q}_i\, \frac{\partial\mathcal{P}}{\partial\dot{q}_i} = k\,\mathcal{P}\chg{.} \label{eq:homogeneous_stationarity}
\end{align}
Contracting the first with $\dot{q}_i$ and using the second together with the balance \eqref{eq:energy_balance} fixes the multiplier away from rest, and with it the kinetic law, that is, the relation between forces and rates, and the unconstrained functional:
\begin{align}
    \lambda = -\frac{k}{k-1}, \qquad X_i = -\frac{1}{k}\, \frac{\partial \mathcal{P}}{\partial \dot{q}_i}, \qquad \mathcal{R} = \dot{\mathcal{F}} + \frac{1}{k}\,\mathcal{P}\chg{.} \label{eq:homogeneous}
\end{align}
For the quadratic form ($k = 2$) these are $\lambda = -2$, the Onsager law\chg{,} and the Rayleighian \eqref{eq:tep_general}. The kinetic law is the one Hackl and Fischer obtain from the principle of maximum dissipation \cite{hackl2008}, $X_i = -\big[\mathcal{P}/\sum_j \dot{q}_j\, \partial\mathcal{P}/\partial\dot{q}_j\big]\, \partial\mathcal{P}/\partial\dot{q}_i$, that is, Ziegler's orthogonality \cite{ziegler1963}; for a homogeneous dissipation its factor is the constant $1/k$, and $\mathcal{P}/k$ is the dissipation potential. For $k \neq 2$ the law is a nonlinear relation between forces and rates, and the remaining statements carry over: $\mathcal{R}$ is convex and minimized by the physical rates, over the balance surface the dissipation is maximal at the same rates, and at a driven stationary state $\mathcal{R}$ reduces to $\mathcal{P}/k$, so that the dissipation itself is minimal over the driving-constrained class.

Both extremal statements refer to equilibrium. For linear kinetics the dissipation is a squared distance: in the metric it defines, $\mathcal{P}$ measures how far the rates (or, through the kinetic law, the forces) are from their equilibrium value, zero. A stationary state driven by prescribed rates \chg{(}boundary velocities or fluxes\chg{)} minimizes $\mathcal{P}$ over the affine class of fields compatible with the driving, and is therefore the point of that class closest to equilibrium: the state as close to equilibrium as the driving permits. It is orthogonal, in the inner product of the dissipation, to the difference of any two fields of the class. This is the weak form of the field equation, the energy orthogonality familiar from Galerkin methods, and it splits the dissipation of any other field of the class into that of the stationary state and that of the difference, as the excess $\mathcal{P} - \mathcal{P}_S$ in phase (a) of Figure~\ref{fig:cavity} illustrates. In a closed inertialess system nothing keeps the state away from equilibrium, and the state descends the free energy along the direction of steepest descent in the same metric. This statement holds at each instant and implies nothing about the duration of the approach. With linear kinetics \chg{and a free energy that is smooth at equilibrium,} equilibrium is never attained during the evolution: the driving forbids it in regime (b) and the approach is asymptotic in regime (c) of Figure~\ref{fig:systems}; for \chg{$k > 2$} it can be reached in finite time. Equilibrium nevertheless organizes both regimes.

\subsection{Implications for coupled problems}
\label{sec:coupled}

The framework carries over directly to coupled multiphysics problems: when several phenomena combine~--- diffusion with mechanics, say, or fluid flow with chemical reactions~--- the coupled Rayleighian functional $\mathcal{R} = \int_V \mathscr{R}\, dV$ is obtained by summing the individual energy rates, dissipation potentials, and constraints. \chg{Four} rules follow: fix each energy rate via the Reynolds transport theorem, choosing the control volume appropriate to the physics; vary only rate fields at the current instant, never applying temporal Euler--Lagrange equations; treat the state variables as given, so that only rates and multipliers are unknowns; \chg{and hold driving forces that contain accelerations fixed during the variation (Sections~\ref{sec:ns} and~\ref{sec:lagrangian})}. The resulting evolution equations are consistent with thermodynamics and free of the ambiguities discussed here.

\section{Conclusion}

We set out to contribute to the understanding of the thermodynamic extremal principle~--- stationarity of the dissipation subject to $\dot{\mathcal{F}} + \mathcal{P} = 0$~--- by investigating its extremal properties. The physical rates \textit{maximize} the dissipation among all balance-consistent rates and simultaneously \textit{minimize} the convex Rayleighian functional $\dot{\mathcal{F}} + \tfrac{1}{2}\mathcal{P}$, which under steady driving reduces to \chg{half the dissipation, $\tfrac{1}{2}\mathcal{P}$} (Figure~\ref{fig:systems})\chg{. All three regimes} appear in a single lid-driven cavity computation (Figure~\ref{fig:cavity}). Whether the TEP \chgb{appears} as a maximum or a minimum is thus decided by the regime; the two families of extremal principles are two \chgb{sides} of one functional.

Methodologically, the TEP is distinct from Hamilton's principle: it optimizes over rates at a fixed instant with purely spatial Euler--Lagrange equations, and for the inertialess Stokes problem the functional restricted to its constraint manifold is exactly quadratic and attains a \textit{global} minimum. The Reynolds transport theorem proved \chg{to be} a useful tool: it supplies the energy rates, and the \chg{kinetic energy rate} it delivers \chg{(}absent, fixed-volume, or material-volume\chg{)} generates the Navier--Stokes family in turn: Stokes, unsteady Stokes, and Navier--Stokes flow. With inertia the driving force \chg{(}$\rho\, D\boldsymbol{u}/Dt$ in Navier--Stokes flow, the Euler--Lagrange expression in Lagrangian mechanics\chg{)} must be held fixed during the variation, less naturally than in the inertialess case\chg{. Writing $\dot{\mathcal{F}}$ as rates times driving forces nevertheless} delivers the Navier--Stokes equation and Lagrange's equations with Rayleigh's dissipation function, so the TEP is not confined to overdamped dynamics. The extremal content survives whenever the frozen force does not depend explicitly on the varied rate\chg{. For} unsteady Stokes flow and constant-mass mechanics the minimum persists as a d'Alembert-type statement, the continuous form of the incremental minimization behind an implicit time step\chg{. For} Navier--Stokes flow the convective term, \chg{nonvariational} in the Eulerian description, ceases to be a \chg{separate} term in material coordinates, where the particle acceleration is the second time derivative of the state and the minimum is recovered without freezing anything by hand\chg{. For} configuration-dependent kinetic metrics only stationarity survives in the given coordinates. The framework extends directly to coupled multiphysics problems (Section~\ref{sec:coupled}).

The examples of this work rest on a dissipation that is quadratic in the rates, that is, on the linear regime of nonequilibrium thermodynamics. This assumption is remarkably often satisfied, holding for Stokes flow, Navier--Stokes flow\chg{,} and Fickian diffusion \cite{flachberger2024} alike, and it restricts only the kinetics: a principle with quadratic dissipation still captures systems that are nonlinear elsewhere, such as Navier--Stokes \chg{flow} (through convective inertia) and dissipative Lagrangian mechanics (through the potential and the kinetic metric). Homogeneity suffices: for a convex dissipation homogeneous of degree $k > 1$ in the rates the Rayleighian becomes $\dot{\mathcal{F}} + \mathcal{P}/k$ and both extremal statements survive (Section~\ref{sec:linearity}). Two questions remain open: the step beyond homogeneity, where the principle of maximum dissipation and the dissipation potential in general part ways \cite{hackl2008}, and the connection to the far-from-equilibrium picture developed over the last decades through \chg{fluctuation relations such as Jarzynski's equality} \cite{jarzynski1997} and England's dissipation-driven adaptation \cite{england2013}.

We close with a thought that goes beyond what has been shown. The most familiar nonequilibrium stationary state of an open system is a living organism: it takes in food and oxygen, produces heat and entropy, and passes matter through, while a working organism holds its temperature, its energy content\chg{,} and its mass approximately constant over time. In the language of Section~\ref{sec:driven} \chg{and Figure~\ref{fig:systems}b} that is $\chg{\dot{\mathcal{U}}} = 0$ and $\dot{\mathcal{F}} = 0$ with $\mathcal{P} = \dot{W}_{\mathrm{ext}}$: the free energy taken in is dissipated entirely, and within the present framework such a state is the constrained \textit{minimum} of the dissipation over the driving-constrained class. The everyday language of biology says the same: an animal in a harsh environment is said to conserve its energy, spending it only on securing the intake that keeps its energy content constant, which in the terms of Section~\ref{sec:driven} is the state of least dissipation compatible with the driving. That homeostasis should be a state of least dissipation is an old intuition: Prigogine and Wiame proposed it in 1946 \cite{prigogine1946}, and the decrease of the specific metabolic rate during development towards an adult plateau \cite{zotin1967} has been read as the approach to it. This would be the biological counterpart of phase (a) in Figure~\ref{fig:cavity}, in which the dissipation decreases monotonically to its stationary minimum. The opposite intuition, that life maximizes entropy production \cite{schneider1994, martyushev2006}, has been defended with equal conviction, and the present work suggests why both can be held at once\chg{. Adaptation} and the irreversible transitions of evolution release stored free energy and belong to the relaxation \chgb{side} of the functional, where dissipation is maximized over the balance-consistent rates\chg{; this} is also the direction in which England's dissipation-driven adaptation points \cite{england2013}\chg{. The} working organism in between belongs to the stationary \chgb{side}, where \chg{dissipation} is minimized over the driving-constrained states. We have shown this only for linear force--flux relations and their homogeneous generalization, and metabolism operates far from that regime; but a relation between homeostasis and the state of minimum dissipation this simple seems to us too fundamental not to be stated, if only as a question.

\end{document}